\documentclass[12pt]{article}
\usepackage{amssymb}
\usepackage{epsfig}
\usepackage{graphicx}
\usepackage{amsmath}
\usepackage{verbatim}

\csname @addtoreset\endcsname{equation}{section}
\begin{document}
\begin{titlepage}
\title{\bf\Large From Supercurrents to Soft Terms \vspace{18pt}}

\author{\normalsize Sibo~Zheng$^{1}$~and~Jia-Hui~Huang$^{2}$ \vspace{12pt}\\
{\it\small  $^{1}$ Department of Physics, Chongqing University, Chongqing 400030, P.R. China}\\
{\it\small $^{2}$ Center of Mathematical Science, Zhejiang University, Hangzhou 310027, P.R. China}\\
}

\date{}
\maketitle \voffset -.3in \vskip 1.cm \centerline{\bf Abstract}
\vskip .3cm In this paper,
 hidden sectors of Ferrara-Zumino multiplets with contributions to
soft terms coming from quantum supergravity are investigated in framework of gravity mediation.
The two-point correlator of Ferrara-Zumino multiplets can be parameterized,
which implies the wave function renormalizations of components fields
in gravity supermultiplet can be evaluated in relatively simple form.
Soft terms are calculated via supercurrent approach.
We find gaugino masses are independent of sfermion masses on general grounds.
The unification of gaugino masses is not universal.
In comparison with general gauge mediation,
there are no sum rules for sfermion masses of each generation.
\vskip 5.cm \noindent August  2010
 \thispagestyle{empty}

\end{titlepage}
\newpage
\section{Introduction}
Among mediation scenarios of supersymmetry breaking,
gravity mediation \cite{CA} is a natural option.
Although gravity mediation suffers from flavor problem that should be taken care of,
it has virtues of achieving locally grand unification and electro weak breaking.
The contributions to soft terms mainly come from anomaly mediation \cite{9810442,9810155}.
In  super-conformal theories, however,
the contributions to soft terms arising from quantum effects of supergravity are dominant.
This paper is devoted to study soft terms that are induced by quantum supergravity.

The complicated Lagrangian of coupling supersymmetric models to supergravity
is the origin of hardly obtaining the structure of soft terms at next leading order.
It is known that the leading order approximation of coupling supersymmetric models to supergravity
can be well described by linearized supergravity (see \cite{Weinberg} and reference therein).
According to the symmetric properties,
the supercurrents ( and corresponding linearized supergravity)
can be classified into Ferrara-Zumion (FZ) multiplet \cite{supercurrents},
$\mathcal{R}$ multiplet \cite{West} and $\mathcal{S}$ multiplet \cite{1002.2228} and
variant supercurrents \cite{Kuzenko,1007.3092}.
In this paper, we will discuss soft terms of FZ multiplet as hidden sector in gravity mediation.
We impose two conditions so that we are able to evaluate the structure of soft terms in relatively simple form.
The first condition is perturbative validity of gravity coupling,
which is always satisfied for soft terms near hundred GeVs.
The other condition is that the FZ multiplet has $R$ symmetry,
which is assumed to break either during embedding the FZ multiplet into supergravity or in the visible sector.
As we will show, even without the second assumption some important results can be still expected.

Starting with FZ multiplet with $R$ symmetry,
we can parameterize the two point correlator of FZ multiplet via a set of functionals.
This procedure is similar to what we have experienced in general gauge mediation \cite{0801.3278}.
We find that gaugino masses are independent on sfermons masses in general and
the unification of gaugino masses is not universal.
The sfermion masses is found to depend on both flavor and gauge quantum numbers,
which implies that there are no sum rule for sfermion masses of each generation.
This property weakens the prediction of gravity mediation at LHC,
however, also separate gravity mediation from gauge mediation.

This paper is organized as follows.
In section 2, we discuss the parametrization of FZ multiplet.
Section 3 is devoted to the calculation of soft terms,
with discussions on phenomenological implications.
Finally, we make a few outlooks.

\section{Supercurrents in Hidden Sector}
In this paper, we follow the conventions of Wess and Bagger \cite{Wess}.
We couple the hidden sector that is responsible for supersymmetry breaking to supergravity via approach of supercurrent.
We consider the supercurrent of hidden sector belongs to the type that can be described by FZ multiplet \cite{supercurrents},
which is viable for a lot of supersymmetric models we are familiar with.
The constraint on supercurrent of FZ multiplet is,
\begin{eqnarray}{\label{constraint}}
\bar{D}^{\dot{\alpha}}J_{\alpha\dot{\alpha}}=D_{\alpha}S=0
\end{eqnarray}
It is understood that $R$ symmetry is a necessary condition for supersymmetry breaking \cite{Seiberg94}.
For hidden sector with this symmetry, $S=0$.
However, note that $R$ symmetry has to be spontaneously broken in order to permit Majorana gaugino mass of visible sector,
we assume this happens when hidden sector embedded into supergravity or in visible sector.

Introducing the two-point correlator of supercurrent,
\begin{eqnarray}{\label{I1}}
<J_{\alpha\dot{\alpha}}(p,\theta)J^{\dot{\alpha}\beta}(-p, \theta')>\equiv~I_{\alpha}^{\beta}(p,\theta,\theta')
=-\frac{1}{2}\sigma^{\mu}_{\alpha\dot{\alpha}}(\bar{\sigma}^{\nu})^{\dot{\alpha}\beta}I_{\mu\nu}(p,\theta,\theta')
\end{eqnarray}
where
\begin{eqnarray}
I_{\mu\nu}(p,\theta,\theta')=<J_{\mu}(p,\theta)J_{\nu}(-p,\theta')>
\end{eqnarray}
From constraint eq\eqref{constraint} we obtain a constraint on $I_{\alpha}^{\beta}$,
\begin{eqnarray}{\label{I2}}
D^{\alpha}D^{\beta}I_{\alpha\beta}=D^{\beta}D^{\alpha}I_{\alpha\beta}=0
\end{eqnarray}
which implies a discrete symmetry $\alpha\rightarrow\beta$,
$p\rightarrow -p$ and $\theta\rightarrow \theta'$ in the strucuture of $I_{\alpha\beta}$.
According to the definition of $I_{\alpha}^{\beta}$ in eq\eqref{I1}, this discrete symmetry requires $I_{\mu\nu}$ is a symmetric tensor.
In this sense, the constraint equation eq\eqref{I2} can be reformulated as,
\begin{eqnarray}{\label{I3}}
D^{2}I(p,\theta,\theta')=0,~~~~~~I=\eta^{\mu\nu}<J_{\mu}J_{\nu}>
\end{eqnarray}
The general solution to eq\eqref{I3} has been considered in literature \cite{0801.3278,Distler},
in which the general form of $I$ was given with four undetermined
scalar functional $F_{(1)}$, $F_{(2)}$, $F_{(3)}$ and $F_{(4)}$.

Roughly there seems to have two options for transforming the scalar expression of the solution to tensor expression.
We can either modify the terms or the coefficients $F_{(i)}$ in front of these terms as required,
or simply extend the coefficients $F_{(i)}$ to tensor functionals $F_{\mu\nu}^{(i)}$.
One can check the later choice is excluded by consistent considerations.
And the simplicity of structure for two point correlator is not kept in general.
However, we can still reduce the complexities to the form that can be handled for FZ multiplets.
As we will see, these parameters contain the information of supersymmetry breaking.

According to the symmetric property, we can write the two point correlator as,
\begin{eqnarray}{\label{correlators}}
<C_{\mu}(p)C_{\nu}(-p)>&=&F^{(3)}_{\mu\nu}(p^{2})\nonumber\\
<\chi_{\mu\alpha}(p)\bar{\chi}_{\nu\dot{\beta}}(-p)>&=&p_{\lambda}\sigma^{\lambda}_{\alpha\dot{\beta}}F^{(2)}_{\mu\nu}(p^{2})
+\left(p_{\mu}\sigma_{\nu}+p_{\nu}\sigma_{\mu}\right)_{\alpha\dot{\beta}}Z(p^{2})\nonumber\\
<\chi_{\mu\alpha}(p)\chi_{\nu\beta}(-p)>&=&\epsilon_{\alpha\beta}M_{P}~F^{(1)}_{\mu\nu}(p^{2})\\
<\hat{T}_{\mu\lambda}(p)\hat{T}_{\nu\kappa}(-p)>&=&-\left(p^{2}\eta_{\lambda\kappa}-p_{\lambda}p_{\kappa}\right)F^{(4)}_{\mu\nu}(p)\nonumber
\end{eqnarray}
Here $F_{(i)}^{\mu\nu}$ and $Z$ are introduced to store the information of supersymmetry breaking.
All the tensor functionals are symmetric. $M_P$ is the mediation scale of supersymmetry breaking.
The last formula in eq\eqref{correlators} is manifested by the conservation of energy-tensor of hidden sector
$\partial^{\mu}\hat{T}_{\mu\nu}=0$.
To derive this formula, we recall the embedding relation of energy-tensor of FZ multiplets,
\begin{eqnarray}
T_{\mu\nu}=-\frac{1}{2}\left[\hat{T}_{\mu\nu}+2g_{\mu\nu}Re(F_{S})\right]=-\frac{1}{2}\hat{T}_{\mu\nu}
\end{eqnarray}
for FZ multiplets with $S=0$.
One can use some simple examples to check the validity of eq\eqref{correlators} for FZ multiplets with $R$ symmetry.
Note that there are no two-point correlators of auxiliary fields $M_{\mu}$ and $N_{\mu}$,
whose contributions to Feynman diagram of soft terms are denoted by their vacuum expectation values (VEV).
Components $\lambda_{\mu}$ and $D_{\mu}$ are unrelated to the calculations of soft terms in this paper,
we do not discuss them.

Coupling the supercurrent of hidden sector to supergravity via
\begin{eqnarray}{\label{hidden}}
\kappa\int d^{4}\theta J_{\mu}H^{\mu}
&=&\frac{\kappa}{2}\left[C_{\mu}D_{H}^{\mu}-(\chi_{\mu}\lambda_{H}^{\mu}+c.c)-\hat{T}_{\mu\nu}\phi_{H}^{\mu\nu}
\right.\nonumber\\
&+&\left.\frac{1}{2}\left((M_{\mu}-iN_{\mu})(M_{H}^{\mu}+iN_{H}^{\mu})+c.c\right)\right]
\end{eqnarray}
from which we obtain the wave function renormalizations of component fields of gravity multiplets.
$\kappa=\sqrt{8\pi~G}$.
According to the normalization taken in eq\eqref{hidden},
all functions $F_{(1)}$, $F_{(2)}$, $F_{(3)}$, and $F_{(4)}$ have mass dimension of two in momentum space.

\section{Soft Terms in Gravity Mediation}
The supercurrent of supersymmetric standard model (SSM) \cite{9506145} is given by ,
\begin{eqnarray}{\label{ssm}}
J^{vis}_{\alpha\dot{\alpha}}=32\bar{W}_{\dot{\alpha}}e^{2V}W_{\alpha}-\frac{2}{3}[D_{\alpha},\bar{D}_{\dot{\alpha}}](Q^{\dag}e^{2V}Q)
+2(\mathcal{D}_{\alpha}Q)~e^{2V}(\bar{\mathcal{D}}_{\dot{\alpha}}Q^{\dag})
\end{eqnarray}
where
$\mathcal{D}_{\alpha}Q=D_{\alpha}Q+(e^{-2V}D_{\alpha}e^{2V})Q$.
$Q_i$ are quark superfields. Coupling the supercurrent of SSM to
supergravity gives in components,
\begin{eqnarray}
\kappa\int d^{4}\theta J_{\mu}H^{\mu}
=\frac{\kappa}{2}\left(C^{J}_{\mu}D_{H}^{\mu}-(\chi^{J}_{\mu}\lambda_{H}^{\mu}+c.c)-\hat{T}^{J}_{\mu\nu}\phi_{H}^{\mu\nu}+M^{J}_{\mu}M_{H}^{\mu}+N^{J}_{\mu}N_{H}^{\mu}\right)
\end{eqnarray}

We divide the Lagrangian into $\mathcal{L}_{F}$ and $\mathcal{L}_{B}$ that are related to
gaugino and sfermion masses respectively, which are explicitly given by,
\begin{eqnarray}{\label{gaugino}}
\frac{1}{\kappa}\mathcal{L}_{F}&=&16D^{\mu}_{H}(\lambda\sigma_{\mu}\bar{\lambda})-\left[16i(\lambda_{H}^{\mu}\sigma^{\nu}\bar{\lambda})\left(F_{\mu\nu}+\tilde{F}_{\mu\nu}\right)
-\frac{8}{3}QQ^{*}(\lambda_{H}^{\mu}\sigma_{\mu}\bar{\lambda})+c.c\right]\nonumber\\
&+&16\phi^{\mu\nu}_{H}\left(i\partial_{\nu}\lambda\sigma_{\mu}\bar{\lambda}+c.c\right)-32\phi^{\mu\nu}_{H}\lambda\sigma_{\mu}\bar{\lambda}V_{\nu}\\
&-&2\left[(M^{\mu}_{H}-iN^{\mu}_{H})\left(4(\sigma^{m}\nabla_{m}\bar{\lambda})\sigma_{\mu}\bar{\lambda}
+\sqrt{2}Q^*\bar{\lambda}\bar{\sigma}_{\mu}\psi\right)+c.c\right]\nonumber
\end{eqnarray}
and
\begin{eqnarray}{\label{sfermion}}
\frac{1}{\kappa}\mathcal{L}_{B}&=&\left[-\frac{4}{3}D_{H}^{\mu}\left(iQ^{*}\partial_{\mu}Q\right)
+2\sqrt{2}(M^{\mu}_{H}-iN^{\mu}_{H})\left(Q^*\psi\sigma_{\mu}\bar{\lambda}\right)+c.c\right]\nonumber\\
&+&\frac{4}{3}\left[\lambda_{H}^{\mu}\left(\sqrt{2}Q^{*}\partial_{\mu}\psi+\sqrt{2}iQ^{*}
V_{m}\sigma_{\mu}\bar{\sigma}^{m}\psi+2QQ^{*}\sigma_{\mu}\bar{\lambda}\right)+c.c\right]\nonumber\\
&-&\frac{1}{2}\phi_{H}^{\mu\nu}\left[\left(\eta_{\mu\nu}Q^{*}\Box~Q-2\eta_{\mu\nu}Q^{*}QV^{2}+4i\eta_{\mu\nu}Q^{*}V^{m}\partial_{m}Q\right)
+c.c\right]\\
&-&2\phi_{H}^{\mu\nu}\left[\left(\eta_{\mu\nu}(D^{m}Q)^{+}(D_{m}Q)+2(D_{\mu}Q)^{+}(D_{\nu}Q)\right)+c.c\right]\nonumber
\end{eqnarray}
where $V_{m}$ and $F_{mn}$ denote the gauge field and field strength
respectively.  $\psi_i$ refer to fermions of standard model and
$Q_i$ their sfermions. The gauge covariant derivative is defined by
$\textrm{D}_{m}Q=\partial_m Q+iV_{m}Q$. $F$ is the auxiliary field
of quark superfield. The terms we neglect are conjugate terms.

We would like to make a few comments about formulas eq\eqref{gaugino} and eq\eqref{sfermion}.
Some of operators in original Lagrangian  do not
contribute to generations of soft masses at leading order, for
example for those terms that only carry derivative of $Q$ scalar
field, thus irrelevant to calculations of sfermion masses.
Some of terms in $\mathcal{L}_{F}$ and
$\mathcal{L}_{B}$ induce soft mass terms at two loop while others at
one loop, which are found to be same order of
$\mathcal{O}(\kappa^{4})$.
Finally, we use components $\lambda_{H}$ and $\phi^{\mu\nu}_{H}$ of gravity supermultiplet
instead of graviton and gravitino fields in evaluating the Feynman diagram. The propagators of $\lambda_{H}$ and $\phi^{\mu\nu}_{H}$
 are obtained via the embedding relations for FZ supermultiplet.
\begin{figure}[!h]
\centering
\begin{minipage}[b]{0.5\textwidth}
\centering
\includegraphics[width=3.5in]{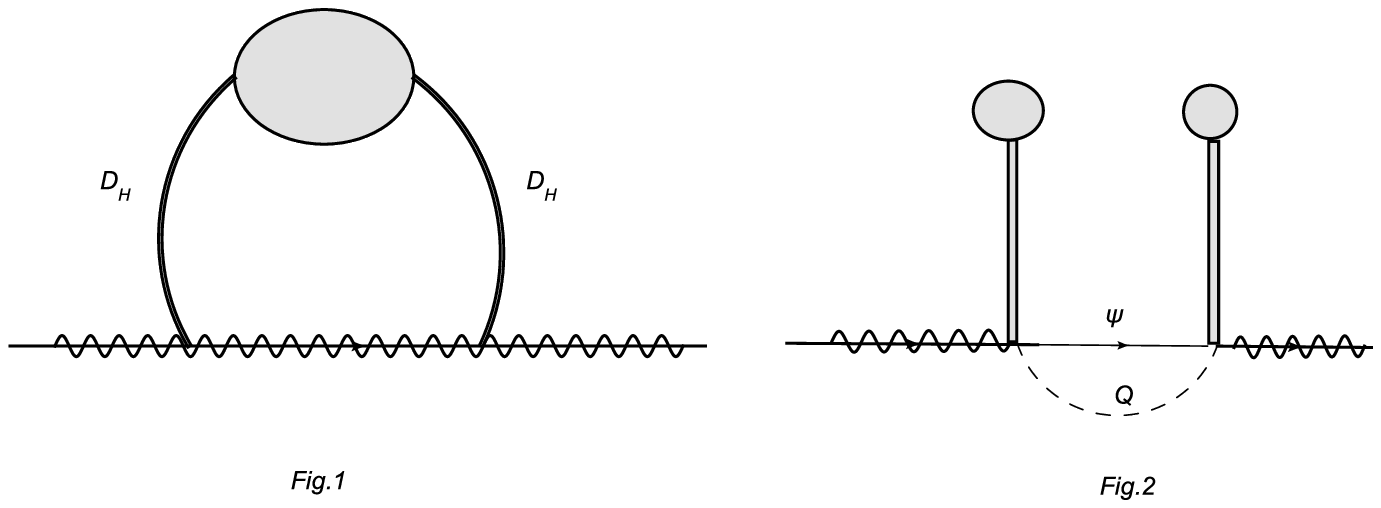}
\end{minipage}%
\begin{minipage}[b]{0.5\textwidth}
\centering
\includegraphics[width=3.5in]{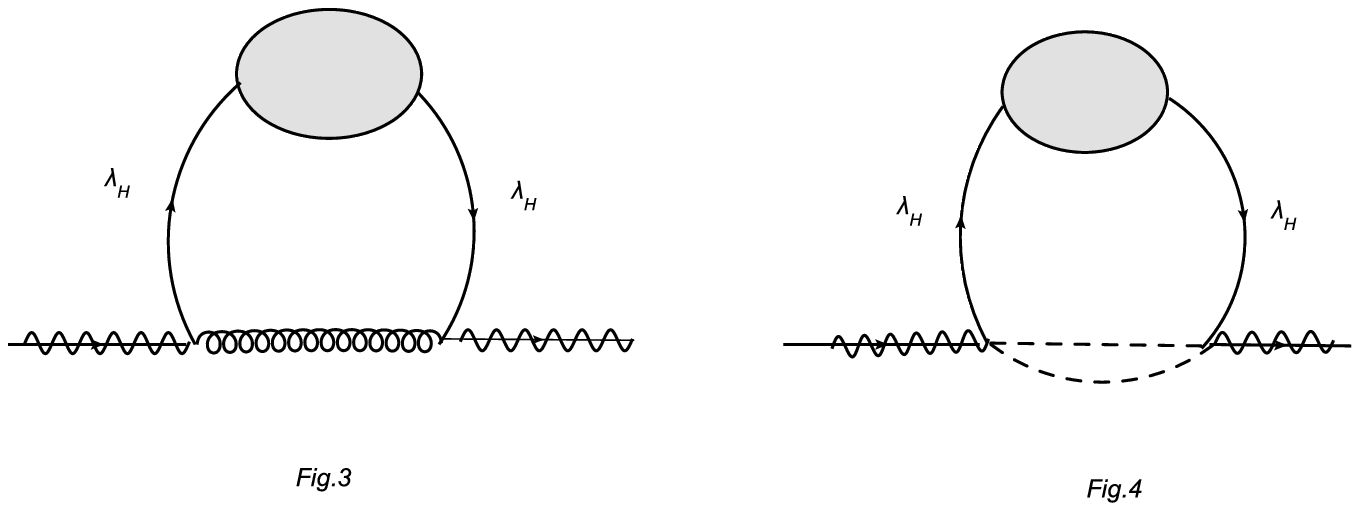}
\end{minipage}
\end{figure}
\begin{figure}[!h]
\centering
\begin{minipage}[b]{0.5\textwidth}
\centering
\includegraphics[width=3.5in]{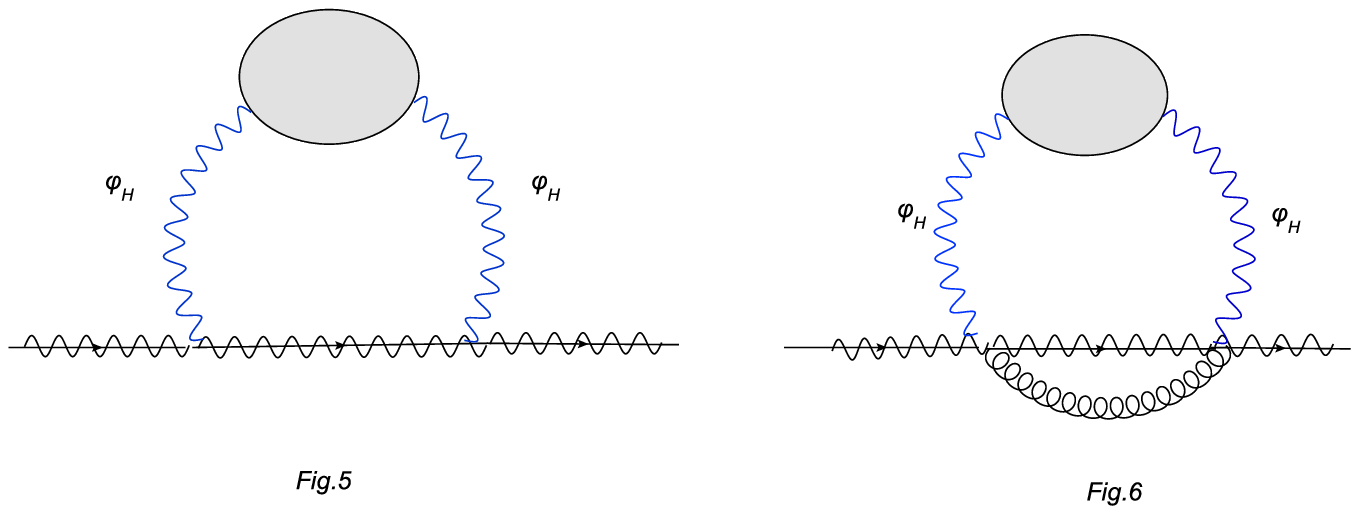}
\end{minipage}%
\begin{minipage}[b]{0.5\textwidth}
\centering
\includegraphics[width=3.5in]{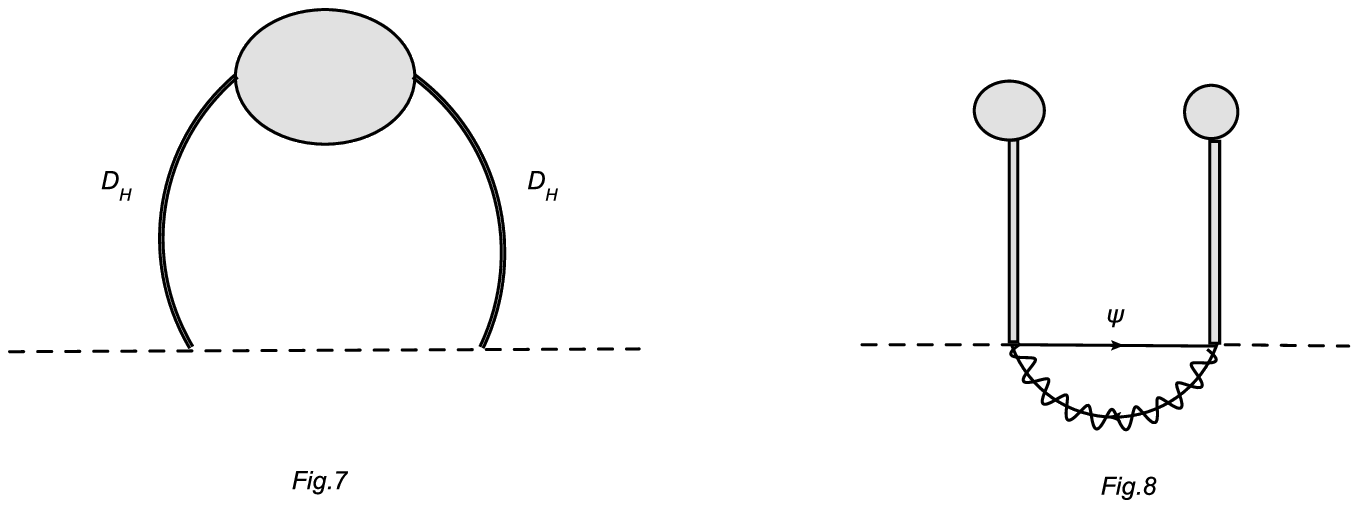}
\end{minipage}
\end{figure}
\begin{figure}[!h]
\centering
\begin{minipage}[b]{0.5\textwidth}
\centering
\includegraphics[width=3.5in]{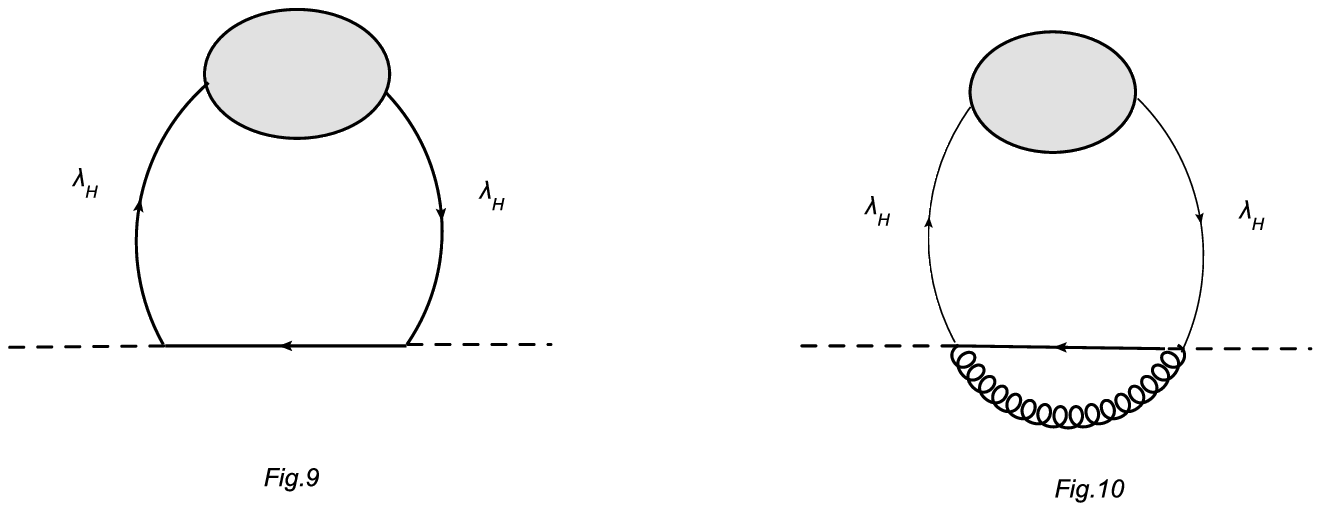}
\end{minipage}%
\begin{minipage}[b]{0.5\textwidth}
\centering
\includegraphics[width=3.5in]{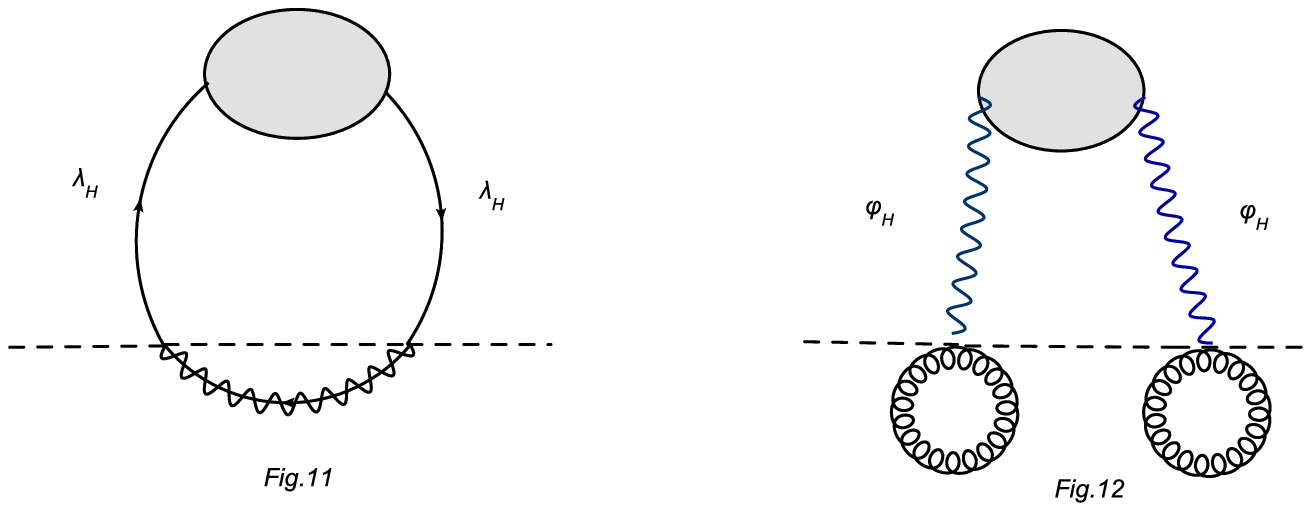}
\end{minipage}
\end{figure}
\begin{figure}[!h]
\centering
\begin{minipage}[b]{0.6\textwidth}
\centering
\includegraphics[width=4in]{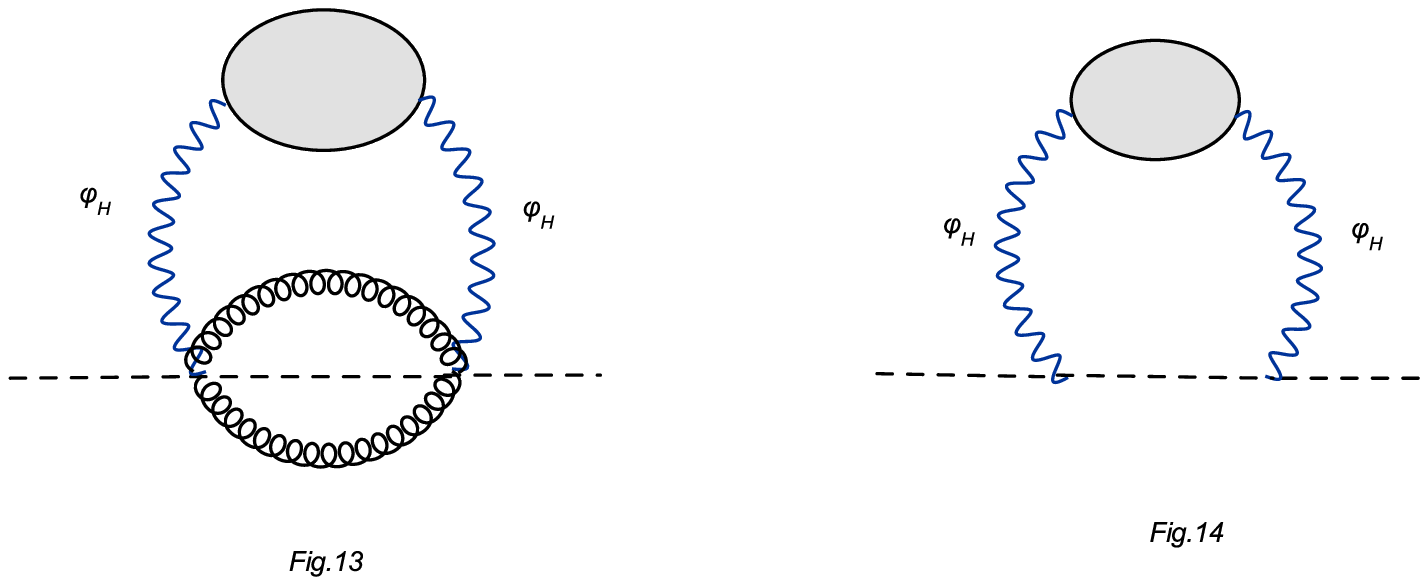}
\end{minipage}%
\begin{minipage}[b]{0.6\textwidth}
\centering
\includegraphics[width=2in]{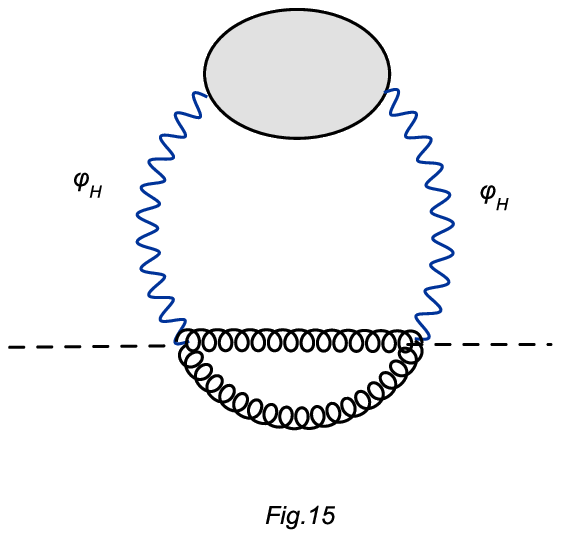}
\end{minipage}
\begin{minipage}[b]{0.2\textwidth}
\centering
\end{minipage}
\end{figure}

According to Lagrangian eq\eqref{gaugino} and eq\eqref{sfermion},
the contributions to gaugino and sfermion masses are composed of Fig.1 to Fig.6 and Fig.7 to Fig. 15 respectively.
First, we computer the gaugino masses.
We find the contributions to gaugino masses only come from Fig. 3 and Fig.4, which are explicitly given by,
\begin{eqnarray}
m_{1/2}
=-\left(2\kappa\right)^{4}\int \frac{d^{4}p}{(2\pi)^{4}}\frac{M_{P}}{p^{4}}\left(L(p)+L^{*}(p)\right)
\end{eqnarray}
where
\begin{eqnarray}
L(p)&=&i2c^{2}_{2}(i)
\left[F_{(1)}^{ab}p_a p_b (308-284i)+F_{(1)}p^2 (418-376i)\right]\nonumber\\
&+&c_{2}(i)\int \frac{d^{4}q}{(2\pi)^{4}}\frac{1}{q^{2}(p-q)^{2}}\left(24p_a p_b F_{(1)}^{ab}+25p^{2}F_{(1)}\right)
\end{eqnarray}
$c_{2}(f,i)$ is the quadratic Casimir of the representation of $f$ under the $\mathrm{r}$ gauge group.

Here are a few comments about gaugino masses.
First, it is observed that there are no contributions coming from those diagram associated with
$F_{(2)}^{\mu\nu}$, $F_{(3)}^{\mu\nu}$ and $F_{(4)}^{\mu\nu}$,
all of which are odd number of powers of momentum integrals.
The gaugino mass is only dependent on function  $F_{(1)}^{\mu\nu}$ via Feynman diagram Fig. 3 and Fig.4.
In comparison with calculations of sfermion mass  as we will show,
the sfermion masses are independent on $F_{(1)}^{\mu\nu}$.
This structure of soft terms has been found to exist in general gauge mediation \cite{0801.3278},
but nerve also expected in gravity mediation.
It can be verified function $F_{(1)}^{\mu\nu}=0$ in supersymmetric limit on general grounds,
which implies that gaugino masses vanish when supersymmetry restored.
Finally, the dependence of gaugino masses on gauge quantum numbers is manifested by $c_{2}(i)$.
Thus, unification of gaugino masses is not universal in gravity mediation. This character is also shared by general gauge mediation \cite{0801.3278}.

Now we compute the sfermion masses.
The contributions to sfermion masses are composed of Fig.7 to Fig.15,
\begin{eqnarray}{\label{sfermion mass}}
m_{0}^{2}&=&-\kappa^{4}\int
\frac{d^{4}p}{(2\pi)^{4}}\frac{1}{p^{2}}\left[\left(\frac{2}{3}\right)^{2}\tilde{K}(p)+\left(\frac{c_{2}(f,i)}{2p^{2}}\right)\tilde{U}(p)
+\left(\frac{4}{3}\right)^{2}\tilde{L}(p)\right.\nonumber\\
&+&\left.p^{2}\left(p^{2}\eta_{\lambda\kappa}-p_{\lambda}p_{\kappa}\right)\tilde{H}^{\lambda\kappa}(p)\right]
\end{eqnarray}
where
\begin{eqnarray}
\tilde{K}(p)&=&p^{\mu}p^{\nu}F^{(3)}_{\mu\nu}\nonumber\\
\tilde{U}(p)&=&-2p^{2}(M^{2}+N^{2})-p_{a}p_{b}\left(M^{a}M^{b}+N^{a}N^{b}\right)
\end{eqnarray}
and $\tilde{L}(p)$ contains the contributions from intermediate
propagator of $\lambda_{H}$, which is function of $F_{(2)}^{\mu\nu}$
and  $Z(p)$.  Since the form of $\tilde{L}(p)$ do not closely
related to main conclusions below, we do not explicitly evaluate it
in this paper. Finally,
\begin{eqnarray}
\tilde{H}^{\lambda\kappa}(p)&=&c^{2}_{2}(f,i)\int\frac{d^{4}k_{1}}{(2\pi)^{4}}
\int\frac{d^{4}k_{2}}{(2\pi)^{4}}\left(\frac{5\cdot(9)^{2}F^{\lambda\kappa}_{(4)}}{p^{4}k_{1}^{2}k_{2}^{2}(p+k_{1}+k_{2})^{2}}
+\frac{9^{2}F^{\lambda\kappa}_{(4)}+F_{(4)}\eta^{\lambda\kappa}}{2p^{4}k_{1}^{2}k_{2}^{2}(p+k_{1}+k_{2})^{2}}\right)\nonumber\\
&+&c_{2}(f,i)\int\frac{d^{4}q}{(2\pi)^{4}}
\left(\frac{2\cdot(9)^{2}F^{\lambda\kappa}_{(4)}}{p^{2}q^{2}(p+q)^{2}}
+\frac{(25p^{2}\eta^{c\lambda}\eta^{e\kappa}
-5p^{\kappa}p^{e}\eta^{c\lambda}-p^{c}p^{\lambda}\eta^{e\kappa}+p^{c}p^{e}\eta^{\lambda\kappa})F_{ce}^{(4)}}{4p^{4}q^{2}(p+q)^{2}}\right)
\nonumber\\
&-&\left(\frac{9}{4}\right)^{2}\frac{1}{p^{2}}F^{\lambda\kappa}_{(4)}+c^{2}_{2}(f,i)\int\frac{d^{4}q}{(2\pi)^{4}}\int\frac{d^{4}k}{(2\pi)^{4}}\left(\frac{5\cdot(9)^{2}F^{\lambda\kappa}_{(4)}}{4p^{4}q^{2}(p+q+k)^{4}}\right)
\end{eqnarray}
We want to mention that in Feynman diagrams associated with $\phi$
scalar fields one should use the intermediate propagator
$<\phi\phi_{\mu\nu}>$.

As mentioned above, sfermion masses do not dependent on $F^{\mu\nu}_{(1)}$,
which is manifested by the Lagrangian eq\eqref{sfermion}.
The vanishing of sfermion masses in supersymmetric limit is not obvious.
Given $F_{(1)}$ and other $F_{(i)}$s of same order,
the gaugino and sfermion masses are roughly comparable with each other. The dependence of sfermion masses on gauge quantum numbers is given by $\tilde{L}(p)$ and $\tilde{H}^{\lambda\kappa}(p)$.
Unlike the universal dependence of sfermon masses on gauge numbers in general gauge mediation \cite{0801.3278, gauge},
the sfermons masses depend on their flavor and gauge symmetries.
Thus, there are no sum rules of sfermion masses of each generation in gravity mediation.
This property weakens the prediction of gravity mediation at LHC, however,
also separates it from gauge mediation.
This helps identifying mediated mechanism of supersymmetry breaking when the primary contributions to soft terms come from quantum supergravity.

There are some interesting issues that should be studied in the future.
First, the positivity of soft terms, especially sfermion masses should be discussed at least in models that are simple enough.
There are some other supercurrent multiplets including
$\mathcal{R}$ multiplets, $\mathcal{S}$ multiplets and variant supercurrents.
It would be interesting to discuss soft terms induced by quantum supergravity
using the supercurrent approach proposed in this paper.

\section*{Acknowledgement}
SZ is supported in part by Chongqing university.
JH is supported by funds from Qiu-Shi, the Fundamental
Research Funds for the Central Universities with contract
number 2009QNA3015.

\end{document}